\documentclass[twocolumn,10pt]{revtex4-1}
\usepackage{amsmath,amssymb}
\usepackage{graphicx}
\usepackage{slashed}
\usepackage{color}

\begin{document}

\title{Momentum-Current Gravitational Multipoles
of Hadrons}

\author{Xiangdong Ji}
\email{xji@umd.edu}
\affiliation{Center for Nuclear Femtography, SURA, 1201 New York Avenue. NW, Washington, DC 20005, USA}
\affiliation{Maryland Center for Fundamental Physics,
Department of Physics, University of Maryland,
College Park, 20742, USA}

\author{Yizhuang Liu}
\email{yizhuang.liu@uj.edu.pl}
\affiliation{Institute of Theoretical Physics, Jagiellonian University, 30-348 Kraków, Poland}

\date{\today}
\begin{abstract}
We study multipole expansion of the momentum currents in hadrons, with three series $S^{(J)}$, $\tilde T^{(J)}$, and $T^{(J)}$, in connection with the gravitational fields generated nearby. The momentum currents are related to their energy-momentum form factors, which in principle can be probed through processes like deeply-virtual Compton scattering currently studied at JLab 12 GeV facility and future Electron Ion Collider. We define the leading momentum-current multipoles, {\it tensor monopole} $\tau$ ($T0$) and {\it scalar quadrupole} $\hat \sigma^{ij}$ ($S2$) moments, relating the former to the so-called $D$-term in the literature. We calculate the momentum current distribution in hydrogen atom and its monopole moment in the basic unit of $\tau_0 =\hbar^2/4M$, showing that the sign of $D$-term has little to do with mechanical stability. The momentum current distribution also strongly modifies the static gravitational field inside hadrons.

\end{abstract}

\maketitle

The energy-momentum tensor (EMT) distribution in a closed system
is an important measure of the underlying dynamics, and according to Einstein's general relativity, is responsible for the space-time geometry nearby. For a stable nuclear system, such as the proton,  deeply-virtual Compton scattering~\cite{Ji:1996nm,HERMES:2001bob,CLAS:2001wjj,ZEUS:2003pwh,H1:2005gdw} and similar processes discussed in
Refs. \cite{Radyushkin:1996ru,Kharzeev:1998bz,Frankfurt:2002ka,Hatta:2018ina,Guo:2021ibg} allows one to measure the EMT form factors, which in the static limit (or disregarding the Compton wavelength) can be interpreted
in terms of the EMT spatial distributions. The EMT effects from long-range forces on the space-time perturbations have been studied in perturbative quantum field theories~\cite{Donoghue:2001qc,Holstein:2006ud}. Here we are interested in bound state systems.

In this paper, we perform the static multipole expansion~\cite{Thorne:1980ru,Blanchet:1998in,Damour:1990gj,Maggiore} of the mass, momentum, and momentum-current (MC) densities in the hadrons of various spins,
trying to understand their physical significance. We relate the hadrons' gravitational form factors to the gravitational multipoles. We find in particular the tensor monopole moment $T0$ is related to the EMT $C(q^2)$ form factor (the  $D$-term in~\cite{Polyakov:1999gs}). As a concrete example of MC distribution, we calculate the $C(q^2)$ form factor in the hydrogen atom, and find that the monopole moment is $\hbar^2/4m$ up to the fractional correction of order $\alpha=1/137$ (fine structure constant). We find
its sign different from the so-called ``mechanical stability" condition~\cite{Polyakov:2018zvc,Lorce:2018egm}. We also remark that the physical significance of ``pressure'' and ``shear pressure"~\cite{Polyakov:2002yz,Polyakov:2018zvc,Burkert:2018bqq,Shanahan:2018nnv} from the momentum current density is only limited to the sense of radiation pressure (see for example~\cite{Noguchi2020}).

Note that the EMT cannot be uniquely derived from
the translational symmetry in the flat space-time. Thus we
always consider matter fields
(electrons, photons, quarks, and gluons, etc)
minimally-coupled to a curved
space-time with metric $g^{\mu\nu}$, and derive the
the energy momentum tensor through
variation~\cite{Callan:1970ze},
\begin{equation}
    T^{\mu\nu} \sim  \frac{\delta S_{\rm matter}}{\delta g^{\mu\nu}} \ ,
\end{equation}
where $S_{\rm matter}$ is the matter field action in the curved space-time.
The flat space-time limit is taken after
the functional derivative.

\section{Static Energy-momentum Tensor multipole expansion}

In this section, we consider multipole expansion of the static EMT distribution $T^{\mu\nu}(\vec{r})$ in a finite system, particularly the momentum current,
$T^{ij}(\vec{r})$. Multipole expansion for electromagnetic systems is well known~\cite{Jackson:1998nia} and we repeat it in the first subsection for the energy current or momentum density $T^{0i}$, which defines two moment series corresponding to two degrees of freedom of the conserved current. The multipole expansion for gravitational systems
has also been worked out in the literature to
considerable details~\cite{Damour:1990gj,Maggiore}. However, most of the studies in gravitational systems focus on generation of gravitational waves. Our interest is in static systems which only involves time-independent moments. We study the moments of $T^{ij}$ in the second subsection, which have three independent series corresponding to three degrees of freedom in the momentum currents. Even though most of the results in this section are not new, they help us to understand the physical significance of the EMT distributions in a quantum mechanical bound state systems such as hadrons or atoms.

We study of the physics of the EMT moments in the context of the linearized Einstein's equation in the weak gravitational limit, in which the symmetric metric tensor $g^{\mu\nu}$ can be approximated by the flat space metric
$\eta^{\mu\nu}=(1,-1,-1,-1)$ (this choice is opposite to the standard convention in the gravitation literature) plus a small perturbation $h^{\mu\nu}$,
\begin{equation}
    g^{\mu\nu} = \eta^{\mu\nu}
  - h^{\mu\nu} \ .
\end{equation}
The rank-2 tensor $h^{\mu\nu}$ has $10$ independent components. Using the coordinates re-parameterization invariance $h_{\mu\nu}\rightarrow h_{\mu\nu}+\partial_{\mu}\zeta_{\nu}+\partial_{\nu}\zeta_{\mu}-\eta_{\mu\nu}\partial^{\rho}\zeta_{\rho}$  where $\zeta_{\mu}$ is an arbitrary vector field, only $6$ components are independent. It is common to define the trace-reversed metric perturbation
\begin{align}
   \bar h^{\mu\nu}=h^{\mu\nu}-\frac{\eta^{\mu\nu}}{2}h^{\rho}_{\rho} \ ,
\end{align}
and a convenient gauge choice is then the harmonic or Lorenz gauge defined by four conditions~\cite{Maggiore}
\begin{equation}
    \partial_\mu \bar h^{\mu\nu} = 0 \ ,
\end{equation}
which are manifestly consistent with the conservation of the EMT.

An important reason to introduce the harmonic gauge is that the trace-reversed metric perturbation satisfies the linearized Einstein's equation
\begin{equation}\label{eq:ein}
    \square \bar h^{\mu\nu} = \frac {16\pi G}{c^4}T^{\mu\nu} \ .
\end{equation}
where $\square = \partial^\mu\partial_\mu$, $G$ is Newton's constant, $c$ the speed of light, and $T^{\mu\nu}$ the EMT of
matter fields. Since Eq.~(\ref{eq:ein}) is just the standard wave equation, it can be solved as~\cite{Maggiore}
 \begin{align}
     \bar h^{\mu\nu}(t,\vec{r})= \frac{4G}{c^4}\int d^3\vec{r}'~ \frac{1}{|\vec{r}-\vec{r}'|}T^{\mu\nu}\left(t-\frac{|\vec{r}-\vec{r}'|}{c},\vec{r}'\right) \ .
 \end{align}
For a time-dependent source, the above will lead to the generation of gravitational waves $(\sim 1/r)$ in regions far away from the source. Among the six independent physical components of the metric, only two correspond to gravitational waves. In the harmonic gauge,  they are defined by the transverse-traceless condition~\cite{Maggiore}. For example, if the wave vector is along the $z$ direction, then the two independent components are $\bar h_{xy}=\bar h_{yx}$ and $\bar h_{xx}=-\bar h_{yy}$.

We will not consider the moment series for the energy/mass density $T^{00}$, which defines,
\begin{equation}
       M_{i_1...i_l} = \int d^3{\vec{r}}~ r_{(i_1}...r_{i_l)}T^{00}(\vec{r})   \ ,
\end{equation}
where $(...)$ are symmetric and trace-free (STF) parts of of the tensor. The generation of gravity by mass multipoles are well-known~\cite{Thorne:1980ru,Blanchet:1998in,Damour:1990gj,Maggiore}.

\subsection{Multipole Expansion for
Energy Current}

Consider the conserved energy current or momentum density $T^{0\mu}$, in a static system. Here for simplicity, we use notation similar to
electromagnetism, with $j^\mu$ standing for $T^{0\mu}$. The static conservation law becomes,
\begin{equation}
    \partial_i j^i(\vec{r}) = 0,
\end{equation}
where $\vec{j}(\vec{r})$ is a static current distribution.  Given the vector current, the vector field $\vec{A}$ (standing for $\bar h^{i0}c^4/(4G)$),
that satisfies the Laplace equation
\begin{align}
    \nabla^2 \vec{A}(\vec{r})=-4\pi\vec{j}(\vec{r}) \ ,
\end{align}
can be solved as
\begin{align}
    A_i(\vec{r})=\int \frac{j_i(\vec{r}') d^3\vec{r}'}{|\vec{r}-\vec{r}'|} \ .
\end{align}
At large distance $r\gg r'$, $A^i$ allows the following multipole expansion
\begin{align}
    A_i(\vec{r})=\sum_{l=0}^{\infty}\sum_{i_1,..i_l}\frac{(-1)^l}{l!}j_{i, i_1...i_l}\partial_{i_1}...\partial_{i_l}\frac{1}{r} \ ,
\end{align}
where the moments of the vector current
\begin{align}\label{eq:jbare}
    j_{i,i_1...i_l}= \int d^3\vec{r} ~r_{(i_1}...r_{i_l)}j_{i}(\vec{r})  \ .
\end{align}
and the symbol $(...)$ again will be used to denotes symmetric and traceless part between tensor indices $i_1$ to $i_l$.  Clearly, the subtraction of trace removes all the moments weighted with $r^2$, $r^4$, etc, which do not yield any new tensor structure, nor contribute to the vector field at large distances.

From group-theoretic point of view, the moments $j_{i,i_1...i_l}$ forms a tensor product of spin-1 and spin-$l$ irreducible representations of the three-dimensional rotation group and can be decomposed into a direct sum of spin-$(l-1)$, $l$ and $l+1$ representations
\begin{align}
    [1]\otimes [l]= [l-1] \oplus [l] \oplus [l+1] \ .
\end{align}
In terms of tensor notation, the above decomposition can be written as~\cite{Damour:1990gj,Maggiore}
\begin{align}\label{eq:momentjde}
    j_{i,i_1...i_l}= U^{(l+1)}_{ii_1...i_l}+\tilde V^{(l)}_{ii_1i_2,...,i_l}+\delta_{i(i_1} V^{(l-1)}_{i_2...i_l)} \ ,
\end{align}
The spin-$l+1$, $l$ and $l-1$ parts $U^{(l+1)}$, $\tilde V^{(l)}$, $ V^{(l-1)}$ reads explicitly
\begin{align}
   &U^{(l+1)}_{ii_1...i_{l}}\equiv j_{(i,i_1...i_l)} \ , \\
   &\tilde V^{(l)}_{ii_1...i_l}\equiv \frac{l}{l+1}j_{[i,i_1]...i_{l}} \ , \\
   &V^{(l-1)}_{i_2..i_{l}} \equiv \frac{2l-1}{2l+1} j_{i,ii_2,...i_{l}} \ .
\end{align}
where indices between $[...]$ are anti-symmetrized.
The above decomposition applies for a generic vector current $\vec{j}$ not necessarily conserved.

For a conserved current, it is easy to show that the totally symmetric $(l+1)$-multipole always vanishes
\begin{align}\label{eq:Uvanish}
    U^{(l+1)}_{ii_1...i_l} \equiv 0  \ ,
\end{align}
or \begin{equation}
    \int d^3\vec{r} ~r_{(i_1}...r_{i_l}j_{i)}(\vec{r})
=0
\end{equation}
which holds with and without trace subtraction. It generates a large number of
identities among the moments after some contractions of indices, for exmaple,
\begin{equation}
    2\int d^3 \vec{r} ~\vec{r}\cdot \vec{j} r_{i_1}
     ...r_{i_{l-1}}
      = -(l-1)\int d^3\vec{r}~r^2 j_{(i_1}r_{i_2}...r_{i_{l-1})} \ .
\end{equation}
For $l=1$, it simply reduces to
\begin{align}\label{eq:G0}
    \int d^3 \vec{r} ~\vec{r}\cdot \vec{j}=0 \ ,
\end{align}
and for $l=2$, the identity reads
\begin{align}\label{eq:G1}
    2\int d^3 \vec{r} ~\vec{r}\cdot \vec{j} r_i=-\int d^3\vec{x} r^2 j_i \ .
\end{align}
These identities will be useful later.

The contribution of the $ V^{(l-1)}_{i_2...i_l}$ multipoles to the vector potential is
\begin{align}
    A_{i}^{V(l-1)}=
    \frac{(-1)^l}{l!} V^{(l-1)}_{i_2...i_l}\partial_{i}\partial_{i_2}...\partial_{i_l}\frac{1}{r}  \ ,
\end{align}
and can be gauged away by the gauge transformation
\begin{align}
    A_{i}\rightarrow A_i-\partial_{i}\bigg(\frac{(-1)^l}{l!}  V^{(l-1)}_{i_2...i_l}\partial_{i_2}...\partial_{i_l} \frac{1}{r}  \bigg)  \ .
\end{align}
Therefore they do not produce a physical effect
in static gauge theories. However, the time-varying $ V$-multipoles are important for time-dependent effects such as radiation, and they form also a useful series for describing the current distribution. The first such a moment is
\begin{align}
  V^{(0)}= \int d^3\vec{r} \vec{r}\cdot \vec{j}(r)=0 \ ,
\end{align}
as a consequence of the identity Eq.~(\ref{eq:G0}).
Thus the first non-vanishing moment appears at
\begin{equation}
 V^{(1)}_i= \frac{3}{5}\int d^3 \vec{r} \left(r_i\vec{r}\cdot \vec{j}(r)-\frac{1}{3}r^2j_i\right)=-\frac{1}{2}\int d^3\vec{r} r^2j_i(\vec{r}) \ .
\end{equation}
In the second equality we have used the identity Eq.~(\ref{eq:G1}). Clearly, this is not independent and relates to the current radius. The independent $ V$-moment series starts from $V^{(2)}_{ij}$, and they can all be related to moments of $\vec{r}\cdot \vec{j}$.

The physically-interesting moment series in the static case are
\begin{align}
    \tilde V_{ii_1...i_{l}}^{(l)}\sim  \int d^3\vec{r}~ m_{i}(\vec{r})r_{(i_1}...r_{i_{l-1)}} \ ,
\end{align}
where the $m_{i}$ is the well-known ``magnetization density''~\cite{Jackson:1998nia} in case of the electric current or ``angular-momentum density'' in case of the energy current $T^i=T^{0i}$,
\begin{align}
    \vec{m}(\vec{r})=\vec{r}\times \vec{j}(\vec{r}) \ , \\
    \vec{J}(\vec{r})=\vec{r}\times \vec{T}(\vec{r})  \ .
\end{align}
$\tilde V^{(1)}$ is just the magnetic moment in electromagnetic and total angular momentum vector $\vec{S}$ for the energy current.

\subsection{Multipole Expansion for
Momentum Currents}

The momentum current $T^{ij}$ can be decomposed into the 3D trace and traceless parts, calling it tensor and scalar parts, respectively. The scalar the multipole expansion defines
\begin{equation}
       S_{i_1...i_l}^{(l)} = \int d^3{\vec{r}}~ r_{(i_1}...r_{i_l)}T_{kk}(\vec{r})  \ ,
\end{equation}
which is similar to the multipoles of the energy/mass density.

For the tensor part, we can make the following mulipole
decomposition~\cite{Blanchet:1985sp,Damour:1990gj,Maggiore},
\begin{align}
    [2]\otimes [l]= [l-2] \oplus [l-1] \oplus [l]
    \oplus [l+1] \oplus [l+2] \ .
\end{align}
In terms of tensor notation, we first define the moments $T_{ij,i_1....i_{l}}$ similar to Eq.~(\ref{eq:jbare})
\begin{align}
    T_{ij,i_1....i_l}=\int d^3\vec{r}~ T_{ij}(\vec{r})r_{(i_1...r_{l})}
\end{align}
where $(i_1...i_l)$ again denotes the traceless and symmetric part. The tensor decomposition then reads~\cite{Blanchet:1985sp,Damour:1990gj,Maggiore}
\begin{align}
    &T_{ij,i_1....i_l}=U^{(l+2)}_{iji_1...i_l}+\tilde U^{(l+1)}_{iji_1....i_l}+\delta_{ii_1} \bar S^{(l)}_{ji_2...i_{l}}\nonumber \\
    &+\delta_{ii_1}\tilde T^{(l-1)}_{ji_2..i_{l}}+\delta_{ii_1}\delta_{ji_2}T^{(l-2)}_{i_3..i_l} \ ,
\end{align}
 where traceless and symmetric subtractions in $ij$ and in $i_1....i_l$ are always assumed.  The multipole series denote~\cite{Blanchet:1985sp,Damour:1990gj,Maggiore}
 \begin{align}
     &U^{(l+2)}_{iji_1...i_l}\equiv T_{(ij,i_1...i_l)} \ , \\
     &\tilde U^{(l+1)}_{iji_1...i_l}\equiv\frac{2l}{l+2}T_{i[j,i_1]...i_l} \ , \\
     &\bar S^{(l)}_{i_1...i_l}\equiv\frac{6l(2l-1)}{(l+1)(2l+3)}T_{i(i_1,i_2....i_l)i} \ ,\\
     &\tilde T^{(l-1)}_{i_1...i_l}\equiv\frac{2(l-1)(2l-1)}{(l+1)(2l+1)}T_{i([i_1,i_{2}]i_3...i_l)i} \ , \\
     &T^{(l-2)}_{i_1....i_{l-2}}\equiv \frac{2l-3}{2l+1}T_{ij,iji_1...i_{l-2}} \ \label{eq:Gl}.
 \end{align}
 where we have included certain coefficients in the definition.

Due to momentum current conservation $\partial_i T^{ij}=0$ in a static system, not all the multipoles above are none-vanishing. One can show that the following general identities are true
\begin{align}\label{eq:generaliden}
   \frac{1}{k!} \sum_{P}\int d^3\vec{r} T_{ii_{P(1)}}r_{i_{P(2)}}....r_{i_{P(k)}}=0 \ ,
\end{align}
where $P$ runs over all permutations $P(1),....P(k)$ of $1,...k$. From the above, one can show that the $l+2$ and $l+1$ moments all vanish
\begin{align}
     U^{l+2}_{iji_1...i_l}\equiv 0 \ ,~~~~ \tilde U^{l+1}_{iji_1...i_l} \equiv 0 \ .
\end{align}
One can form more identities from Eq.~(\ref{eq:generaliden}) by performing contractions or symetrization/antisymetrizations.
For example, by contracting $i$ with one of the other indices under permutation one has
\begin{align}
    \int d^3\vec{r} T_{ii}r_{(i_1}...r_{i_k)}=-k\int d^3\vec{r} r_iT_{i(i_1}r_{i_2}....r_{i_k)} \ .
\end{align}
By contracting two of the indices under permutation, one has
\begin{align}
    2\int d^3\vec{r}T_{ij}r_{j}r_{i_1}....r_{i_{k-2}}=-(k-2)\int d^3\vec{r} r^2T_{i(i_1}r_{i_2}...r_{i_{k-2})} \ ,
\end{align}
and so on. For $k=2$ and $k=3$, the above reduces to
\begin{align}\label{eq:k=2}
    \int d^3\vec{r} r_iT_{i(i_1}r_{i_2)}=-\frac{1}{2}\int d^3\vec{r} T^{ii}r_{(i_1}r_{i_2)} \ ,
\end{align}
and
\begin{align}\label{eq:k=3}
    \int d^3\vec{x} r^2 T_{(i_1i_2)}=-2\int d^3\vec{r} r_iT_{i(i_1}r_{i_2)}=\int d^3\vec{r} T^{ii}r_{(i_1}r_{i_2)} \ .
\end{align}
Notice that above holds without trace subtraction as well.

Moreover, by performing anti-symetrization of $i$ with one of the indices under permutation in Eq.~(\ref{eq:generaliden}) one obtain
\begin{equation}
    \int d^3\vec{r} T_{i[j}r_{k]} = 0  \ ,
\end{equation}
and
\begin{align}\label{eq:G1=0}
    \int d^3\vec{r} (r_{[j}T_{k]i_1} r_{i_2} +r_{[j}T_{k]i_2}r_{i_1})= 0 \ .
\end{align}
and so on.

Given theses relations, we can re-express the tensor momentum current multipole $\bar S^{(l)}_{i_1...i_l}$ in terms of that of scalar momentum current multipoles. First let's consider $l=2$, in this case the above reduces to
\begin{align}
   &\bar S^{(2)}_{i_1i_2}=\frac{12}{7}\int d^3\vec{r} \nonumber \\
   &\times \bigg( r_i T_{i(i_1}r_{i_2)}-\frac{1}{3}r^2T_{(i_1i_2)}-\frac{1}{3}T_{ii}r_{(i_1}r_{i_2)}\bigg)
\end{align}
which by using Eq.~(\ref{eq:k=2}) and Eq.~(\ref{eq:k=3}) reduces to
\begin{align}
 \bar S^{(2)}_{i_1i_2}= -2\int d^3\vec{r} r_{(i_1}r_{i_2)} T_{ii}(\vec{r})\equiv -2\sigma_{i_1i_2} \ ,
\end{align}
where the quadrupole of the scalar momentum current $\sigma_{i_1i_2}$ or {\it scalar quadrupole} $S2$ is defined as
\begin{align}
    \sigma_{ij}\equiv S^{(2)}_{ij}=\int d^3\vec{r} T_{kk}(\vec{r})r_{(i}r_{j)} \ .
\end{align}
In fact, for general $l$ one can show that the above relation remains valid~\cite{Damour:1990gj}
\begin{align}\label{eq:Trelation}
  \bar   S^{(l)}_{i_1i_2,....i_l}=-2S^{(l)}_{i_1...i_l} \ ,
\end{align}
therefore, at given order $l$, there is only one series of linear independent spin-$l$ multipoles $S^{(l)}_{i_1...i_l}$.

Given the moments, one can study their contribution to $\bar h^{ij}$. By standard methods, the contribution of $\bar S$ and $S$ reads~\cite{Damour:1990gj,Maggiore}
\begin{align}
    \bar h^{Sl}_{ij}=&\frac{4G(-1)^l}{l!}\nonumber \\
    \times \bigg(&\frac{\delta_{ij}}{3}(S^{(l)}_{i_1..i_l}-\bar S^{(l)}_{i_1..i_l})\partial_{i_1}..\partial_{i_l}\frac{1}{r}+\bar S^{(l)}_{i_1...i_{l-1}i}\partial_{j}\partial_{i_1}..\partial_{i_{l-1}}\frac{1}{r}\bigg) \ ,
\end{align}
where symetrization between $i$ and $j$ is assumed. Using the relation Eq.~(\ref{eq:Trelation}), the above can be written in the form
\begin{align}
   \bar h^{Sl}_{ij}= \delta_{ij}\partial_{k}\zeta_k-\partial_{i}\zeta_j-\partial_j\zeta_j
\end{align}
where
\begin{align}
    \zeta_{i}^{Sl}=\frac{4G(-1)^l}{l!}S^{(l)}_{kk,ii_1....i_{l-1}}\partial_{i_1}..\partial_{i_{l-1}}\frac{1}{r} \ ,
\end{align}
therefore, after a gauge transformation
\begin{align}
    \bar h_{\mu\nu}\rightarrow \bar h_{\mu\nu}+\partial_{\mu}\zeta_\nu+\partial_\mu\zeta_\nu-\eta_{\mu\nu}\partial^{\alpha}\zeta_{\alpha} \ ,
\end{align}
we are left only with the contribution in $\bar h^{00}$~\cite{Damour:1990gj,Maggiore}
\begin{align}
    \bar h_{00}^{Sl}=\frac{4G(-1)^l}{l!}S^{(l)}_{i_1..i_l}\partial_{i_1}..\partial_{i_l}\frac{1}{r} \ ,
\end{align}
which is just the standard scalar multipole expansion with $T_{kk}(\vec{r})$ as the scalar density, similar
to the energy density multipoles. The leading contribution
comes from the scalar quadrupole $S2$ moment.

Finally we come to the other two $\tilde T^{(l-1)}$ and $T^{(l-2)}$ multipole series. Their contribution to the trace-reversed metric perturbation $\bar h^{ij}$ reads~\cite{Damour:1990gj,Maggiore}
\begin{align}
    \bar h_{ij}^{\tilde T(l-1)}=\frac{2G(-1)^l}{l!}\tilde T^{(l-1)}_{i_1ii_2...i_{l-1}}\partial_{j}\partial_{i_1}...\partial_{i_{l-1}}\frac{1}{r}+(i\rightarrow j) \ ,
\end{align}
 and
 \begin{align}
     \bar h^{T(l-2)}_{ij}=\frac{4G(-1)^{l}}{l!}T^{(l-2)}_{i_2...i_{l}}\partial_{i}\partial_{j}\partial_{i_2}...\partial_{i_l}\frac{1}{r} \ .
 \end{align}
Similar to the case of the vector current, they can all be gauged away through gauge transformations
\begin{align}
    \zeta^{\tilde T(l-1)}_i=\frac{2G(-1)^l}{l!}\tilde T^{(l-1)}_{i_1ii_2...i_{l-1}}\partial_{i_1}...\partial_{i_{l-1}}\frac{1}{r}\ ,
\end{align}
 and
 \begin{align}
     \zeta^{T(l-2)}_i=\frac{2G(-1)^{l}}{l!}T^{(l-2)}_{i_2...i_{l}}\partial_{i}\partial_{i_2}...\partial_{i_l}\frac{1}{r} \ .
 \end{align}
 Since $\partial^i \zeta^{\tilde T(l-1)}_i=\partial^i\zeta^{T(l-2)}_i=0$ due to antisymetrization in $T^{(l-2)}$ and $\partial^2\frac{1}{r}=0$ at large distance, the gauge transformation will not produce new terms in $\bar h^{00}$. Therefore, both series $\tilde T^{(l-1)}$ and
$T^{(l-2)}$ have no physical effect at large distance in static case.

However, moment series $\tilde T^{(l-1)}$ and $T^{(l-2)}$ still provide useful characterization of the momentum current distribution (and
do have physical effects in time-varying systems). At $l=1$, the $\tilde T^{(0)}$ vanishes. At $l=2$, non-vanishing moment is related to the tensor momentum-current monopole or {\it tensor monopole} $T0$ for short:
 \begin{align}
     T^{(0)}& =\frac{1}{5}\int d^3\vec{r}T_{ij}(\vec{r})\left(r_ir_j-\frac{\delta_{ij}}{3}r^2\right) \nonumber \\
     & \equiv \frac{2}{15}\int d^3\vec{r}
     r^2s(r)\ ,
 \end{align}
where the second line serves as
a definition for $s(r)$ ("shear pressure").
Using Eq.~(\ref{eq:k=2}), it is related to {\it scalar momentum-current radius} ,
\begin{align}
    T^{(0)} &=-\frac{1}{6}\int d^3\vec{r} r^2 T_{ii}(\vec{r})
    \nonumber \\
    & = - \frac{1}{2} \int d^3\vec{r} r^2 p(\vec{r})  \ ,
\end{align}
where $T^{ii}$ is proportional to the so-called pressure $p(r)$ in the other literature~\cite{Polyakov:2002yz,Polyakov:2018zvc,Burkert:2018bqq,Shanahan:2018nnv}. We choose to define the
tensor-MC monopole moment of a system as
\begin{equation}
    \tau =-T^{(0)}/2 \ ,
\end{equation}
which relates to the ``$D$-term'' $D(0)$~\cite{Polyakov:1999gs} as $\tau=\frac{D(0)}{4M}$.

We next come to the $\tilde T1$ dipole $\tilde T^{(1)}_{ij}$. It is anti-symmetric in $ij$ can be written as
\begin{align}
    \tilde T^{(1)}_{ij}=\frac{2}{5}\int d^3\vec{r}r_kT_{k[i}(\vec{r})r_{j]}\ .
\end{align}
If one define the dilatation current at $t=0$,
\begin{align}
    j^i_D=r_k T^{ki} \ ,
\end{align}
then $\tilde T^{(1)}$ can be conveniently expressed as the ``magnetic moment'' of the dilatation current.  Due to Eq.~(\ref{eq:G1=0}), $\tilde T^{(1)}_{ij}=0$ identically.

To summarize, there are three series of multipoles
for momentum current, $S^{(l)}$, $\tilde T^{(l-1)}$, and $T^{(l-2)}$.
The leading-order moments are tensor monopole $T0$ $\tau=-T^{(0)}/2$, and scalar quadrupole $S2$ $\sigma_{ij}=S^{(2)}_{ij}$, with vanishing
tensor dipole $\tilde T1$. To the next order,
one has scalar octupole $S3$, tensor quadrupoles $\tilde T2$ and $T2$,  and tensor dipole $T1$, and so on.

\section{EMT form factors of hadrons and
Gravitational multipoles}

In this section, we consider examples of the gravitational multipoles in hadrons of different spin.
Not all hadrons are capable of generating all types of
gravitation multipoles. For spin-0 particle such as the pion or $^4$He nucleus, only two multipoles can be generated, one corresponds to the total mass $M$ (mass monopole $M0$), and the other to the momentum-current tensor monopole $\tau$ ($T0$). For a spin-1/2 hadron such as the proton and neutron, one can generate in addition the angular-momentum dipole ($\tilde V1$).
For a spin-1 resonance, such as $\rho $ meson,
one can generate the mass quadrupole $M2$, scalar quadrupole $S2$, and
tensor quadrupole $T2$. In the following we will discuss each of them in turn.

We work in the limit that the hadron masses are large so that their Compton wavelength is negligible~\cite{Jaffe:2020ebz}. This is true in the large $N_c$ limit for baryons (certainly not true for a pion). In this case,
one can directly Fourier-transform the form factors to the position space to obtain the space density distributions. For particles whose
Compton wavelength is not small compared
to its size, an option is to go to the infinite momentum frame~\cite{Burkardt:2002hr,Miller:2007uy,Freese:2021mzg} where one has to be content with
a 2D interpretation. In practice, we adopt the standard Breit frame
approach as the {\it definition} of a spatial density. When
studying the gravitational perturbation at the distance $r$ much larger than the Compton wave length, the formula in terms of the form factors are accurate, independent of the density
interpretation.

\subsection{Spin-0 case}

Let's first consider a scalar system. The EMT
matrix element between the plane wave states $|P^\mu\rangle$
and $|P^{\mu'}\rangle$ defines the
gravitational form factors, $A$ and $C$~\cite{Donoghue:2001qc,Polyakov:2018zvc}
\begin{align}
    &\langle P'|T^{\mu\nu}|P\rangle
    \nonumber \\ &= 2P^{\mu}P^{\nu}A(q^2) +2(q^{\mu}q^{\nu}-g^{\mu\nu}q^2)C(q^2) \ ,
\end{align}
where $q^\mu=P'^\mu-P^\mu$ is the momentum transfer.
The momentum conservation requires $A(0)=1$.
In the Breit frame where $\vec{P}+\vec{P}'=0$,
the Fourier transformation of $MA(q)$ corresponds
to a part of the mass density $\rho_m(\vec{r})$.

The form factor $C(q^2)$ is related to the tensor-MC monopole distribution
in the system, besides contributing to
the mass density $\rho_m(\vec{r})$. In fact, if we Fourier-transform $T^{ij}$ to the coordinate space, it has
the form
\begin{align}
    T^{ij}(\vec{r})=(\nabla^2\delta^{ij}-\nabla^i\nabla^j)\frac{C(r)}{M} \ .
\end{align}
where $C(r)$ is the Fourier version
of $C(q^2)$ and we have divided a factor $2M$ from the relativistic normalization ($\langle P|P'\rangle =
(2\pi)^32P^0\delta^3(\vec{P}-\vec{P}')$).  A simple calculation shows
that the MC monopole moment $T0$ is just,
\begin{equation}
         \tau = \frac{C(q=0)}{M} \ ,
\end{equation}
which relates to the $D$-term~\cite{Polyakov:1999gs} as $\tau=\frac{D(0)}{4M}$. For a free spin-0 boson
~\cite{Donoghue:2001qc},
\begin{equation}
  \tau_{\rm boson} = -\frac{\hbar^2}{4M}
\end{equation}
when proper SI unit ${\rm kg}\cdot {\rm m}^4 \cdot {\rm s}^{-2}$ is restored. We define a fundamental unit $\tau_0 = \frac{\hbar^2}{4M}$, and write $\tau_{\rm boson} = g_b\tau_0$, then $g_b=-1$. For a minimally-coupled interacting theory,
the monopole moment remains the same~\cite{Callan:1970ze}. For more complicated examples including non-minimal coupling, QCD pseudo-Goldstone bosons, as well as large nuclei, see Refs. \cite{Callan:1970ze,Hudson:2017xug,Polyakov:2018zvc} for extensive discussions. Monopole density distribution is related to $s(r)$ defined
in Eq. (60).

For a system with long-range force, such as a charged
particle, it can be shown that $C(\vec{q}\rightarrow 0)$ is infrared divergent~\cite{Berends:1975ah,Milton:1976jr,Kubis:1999db,Donoghue:2001qc,Varma:2020crx,Metz:2021lqv}:
\begin{align}
    \frac{C(\vec{q})}{2M}\rightarrow \frac{\alpha \pi }{16|\vec{q}|}+\frac{\alpha}{6\pi M} \ln \frac{\vec{q}^2}{M^2} \ .
\end{align}
The first term is due to the large $r$ asymptotic decay of Coulomb potential and is classical in nature~\cite{Donoghue:2001qc,Kubis:1999db}, while the second term is quantum in nature. It can be shown~\cite{Donoghue:2001qc} that a divergent monopole moment $C(0)$ ($\tau_{\rm eff}=+\infty$) will generate $1/r^2$ correction to the space-time metric
\begin{align}\label{eq:spin0p}
    \bar h^{ij}(\vec{r})=\frac{G\alpha \hat r_i\hat r_j}{r^2} +\frac{4G\alpha}{3\pi Mr^3}(\hat r^i\hat r^j-\delta^{ij}) \ ,
\end{align}
where the first term comes form the linear divergent part $\frac{\pi \alpha}{16|q|}$ of $C(q)$, while the second term is due to the logarithmic divergent term $\frac{\alpha}{6\pi M} \ln \frac{\vec{q}^2}{M^2}$.

\subsection{$C(q^2)$ Contribution
to Gravitational Potential}

According to Sec. I, it appears that the tensor monopole $\tau$ does not contribute to the long-distance properties of the gravity, other than
it produces a pure gauge contribution. However,
form factor $C(q^2)$ does generate a short-distance static gravitational
potential $h^{00}$ through its contribution to the energy density.

It can shown by solving the linearized Einstein equation that
\begin{align}\label{eq:hC}
     h^{00}_C(\vec{r}) &= -\frac{8\pi G}{c^4M} C(r) \ , \\
     h^{ij}_C(\vec{r})&=    \frac{8\pi G}{c^4M}  C(r)  \delta^{ij} \ ,
\end{align}
thus $C(r)$ contributes a part of the gravitational potential
at short distance! The total
static potential will be added by
the form factor $A(q^2)$ contribution
\begin{align}
     h^{00}_A(\vec{r}) &= \frac{2}{c^4} V(r) \ , \\
     h^{ij}_A(\vec{r})&=   \frac{2}{c^4}  V(r)  \delta^{ij} \ ,
\end{align}
where
\begin{align}
    V(r)=\int d^3\vec{r}'\frac{GM}{|\vec{r}-\vec{r}'|} A(\vec{r}') \ .
\end{align}
At large $r$, $h_C^{\mu\nu}$ decays exponentially whereas
$h_A^{\mu\nu}$ reduces to a point-mass Newton's potential.
For a non-relativistic probe, only $h^{00}$ matters.

\subsection{Spin-$\frac{1}{2}$ case}

For a spin-1/2 system, the matrix element of the EMT is~\cite{Kobzarev:1962wt,Pagels:1966zza,Ji:1996ek}
\begin{align}
&\left\langle P^{\prime}\left|T^{\mu \nu}\right| P\right\rangle=\bar{u}\left(P^{\prime}S'\right)\Bigg{[}A(q^2) \gamma^{(\mu} \bar{P}^{\nu)}
\nonumber \\& +B(q^2) \frac{\bar{P}^{(\mu} i \sigma^{\nu) \alpha} q_{\alpha}}{2 M} +C(q^2) \frac{q^{\mu} q^{\nu}-g^{\mu \nu} q^{2}}{M}\Bigg{]} u(PS) \ ,
\end{align}
where $\bar P^\mu = (P + P')^\mu /2$.
There are now three dimensionless gravitational form factors $A(q^2)$, $B(q^2)$ and $C(q^2)$. The physics of $A(q^2)$ is
the same as that for the spin-0 case.

The form factor $B(q^2)$ is related to the angular momentum
distribution in the system. Indeed, the momentum density $T^{0i}$ is
\begin{align}
    \vec{T}(\vec{r})=-\frac{1}{2}\vec{S}\times \nabla \big(A(r)+B(r)\big) \ .
\end{align}
where $A(r)$ and $B(r)$ are Fourier transformation
of $A$ and $B$ form factors, which
generates the new $\bar h^{i0}$ perturbation through
the angular momentum density $\vec{S}(\vec{r}) =  \vec{r}\times \vec{T}(\vec{r})$. The $\tilde V1$ moment of the momentum density
yields total angular momentum $\vec{S}$. Angular momentum
conservation requires $B(q^2=0)=0$.

The MC monopole moment is zero
for a free fermion~\cite{Hudson:2017oul}.
In general, a spin-1/2 system has
\begin{equation}
      \tau = \frac{C(0)}{M} \ .
\end{equation}
There have been extensive studies in the literature about
$C(0)$ for the nucleon~\cite{Polyakov:2018zvc, Neubelt:2019sou}. In particular, lattice QCD calculations have been made for the separate contributions from quarks and gluons~\cite{LHPC:2007blg, Shanahan:2018nnv}. It appears that $\tau_N$ is negative from various nucleon models as well as preliminary lattice result.

\subsection{Spin-$1$ case}

Unlike the spin-0 and spin-$\frac{1}{2}$ cases, a spin-1 hadron has $6$ independent gravitational form factors~\cite{Holstein:2006ud,Taneja:2011sy,Cosyn:2019aio}
\begin{align}
    &\langle P', \epsilon_f|T^{\mu\nu}(0)|P, \epsilon_i\rangle \nonumber \\ &
    = -2\bar P^{\mu}\bar P^{\nu} \left[(\epsilon^{\star}_f\cdot \epsilon_i) A(q^2) + E^{\alpha\beta}q_\alpha q_\beta \frac{\tilde A(q^2)}{M^2}\right] \nonumber \\ & +J(q^2) \frac{i\bar P^{(\mu}S^{\nu)\alpha}q_\alpha}{M}  &\nonumber\\ &- 2(q_{\mu}q_{\nu}-g_{\mu\nu}q^2) \left[(\epsilon^{\star}_f\cdot \epsilon_i)C(q^2)    + E^{\alpha\beta}q_\alpha q_\beta \frac{\tilde C(q^2)}{M^2}\right] \nonumber \\ &-\left[(E^{\mu\nu}q^2- E^{\mu\alpha}q^\nu q_\alpha -
    E^{\alpha\nu}q^\mu q_\alpha + g^{\mu\nu} E^{\alpha\beta}q_\alpha q_\beta \right]D(q^2)  \ .
\end{align}
where $\epsilon_i$ and $\epsilon_f^*$ are polarization four-vectors of the initial and final
hadron, satisfying $\epsilon_i \cdot P=\epsilon_f \cdot P' =0$. To simplify the expression, we also use symmetric polarization
density matrix
\begin{equation}
E^{\mu\nu}= \frac{1}{2}\left(\epsilon^{*\mu}_f\epsilon^\nu_i
+\epsilon^{*\nu}_f\epsilon^\mu_i\right) \ ,
\end{equation}
and the anti-symmetric polarization tensor
\begin{equation}
  S^{\mu\nu} = i(\epsilon^{*\mu}_f\epsilon^\nu_i
-\epsilon^{*\nu}_f\epsilon^\mu_i) \ .
\end{equation}
For the convenience of calculating multipoles, the above definition of the dimensionless form factors is slightly different from other conventions. An example of $\rho$
meson EMT form factors can be found in~\cite{Epelbaum:2021ahi} using chiral perturbation theory and in~\cite{Fress2019} from Nambu-Jona-Lasinio model calculation.

To relate the above form factors to gravitational multipoles,
we consider the static limit where initial and final states have small momenta, and $\epsilon^0=0$ and $\vec{\epsilon}_i=\vec{\epsilon}_f=\vec{\epsilon}$ is arbitrary. Looking at the $T^{00}$, $A(q^2=0)=1$ gives rise to the mass-monople contribution, while $\tilde A(q^2=0)$ contributes to the mass quadrupole $M2$,
\begin{equation}
M_{ij} =\int d^3\vec{r}\left(r_ir_j-\frac{1}{3}\delta_{ij}r^2\right)\rho_m(\vec{r})
=\frac{2\tilde A(0)}{M}\hat E_{ij}
\end{equation}
where $\hat E^{ij}$ is the traceless part of  $E^{ij}$. The mass quadrupole generates $1/r^3$ perturbation in $\bar h^{00}$ in the following form
\begin{align}
     \bar h^{00}&=\frac{2G M_{ij}}{r^3} (3\hat r^i\hat r^j-\delta^{ij})\ .
\end{align}
The momentum density $T^{0i}$ in Fourier space is,
\begin{align}
    T^{0i}=\frac{i}{2} S^{ij}q_jJ(q^2)=\frac{i}{2}(\vec{S}\times q)^i J(q^2)\ ,
\end{align}
where the axial vector $\vec{S}={\rm Re}(i\vec{\epsilon}^{\star}\times \vec{\epsilon}) $, from which one identify $J(\vec{r})$ as the angular momentum dipole density. Angular momentum conservation constraints $J(q^2=0)=1$.

From the expression for $T^{ij}$, one reads off the tensor-MC monopole $T0$ moment
\begin{equation}
    \tau = \frac{C(0)}{M} \ ,
\end{equation}
which is zero for a free photon.
The monopole moment of the $\rho$ messon appears close to that of the pion. For other spin-1 systems including deuteron, see Ref.\cite{Polyakov:2018zvc}.

There is also a new tensor-MC quadrupole $T2$ moment
\begin{equation}
        T^{(2)}_{ij}=-\frac{\tilde C(0)}{48M^2}\hat E_{ij}
\end{equation}
where the multiple series $T^{(2)}_{ij}$
is defined in Eq.~(\ref{eq:Gl}). The contribution $\tilde C(0)$ to the scalar momentum current is new for spin-1 systems.

Finally, the scalar-MC quadrupole moment $S2$ can be calculated as,
\begin{equation}
\sigma_{ij} = \frac{D(q^2=0)}{M}\hat E_{ij} \ ,
\end{equation}
Thus the tensor quardupole is proportional to $D(0)$ form factor defined above. After gauge transformation, it will generate contribution to $\bar h^{00}$ as
\begin{align}
    \bar h^{\tau,00}&=\frac{2G \sigma_{ij}}{r^3} (3\hat r^i\hat r^j-\delta^{ij}) \ ,
\end{align}
which can be combined with the one from mass-quadruple into the form
\begin{align}
\bar h^{00}=\frac{2G (M_{ij}+\sigma_{ij})}{r^3} (3\hat r^i\hat r^j-\delta^{ij}) \ ,
\end{align}
in consistent with the general results in Refs.~\cite{Damour:1990gj,Maggiore}.

\section{Scalar Momentum-Current
Distribution and $T0$ Moment in Hydrogen Atom}

In this section we study the EMT of the hydrogen-like atom. Contrary to the single charged electron, hydrogen-like atoms are charge neutral and are expected to have finite scalar-MC monopole moment $\tau$. We first show that in the quantum mechanics it is possible to construct a conserved EMT using quantum-mechanical wave functions. However, it still has long-range Coulomb tail due to the interaction between the electron and the proton. We then show that the above conserved EMT can be justified in the field theoretic framework and identified as the leading-order electron kinetic contribution plus the leading order Coulomb photon exchange contribution. By adding the single electron and single proton contributions, the Coulomb tail gets removed and the resulting monopole moment $\tau$ is equal to the basic unit
$\tau_0=\hbar^2/4M$ and {\it positive}. We argue that the result is accurate to leading order in $\alpha$. This example shows that the sign of the EMT has little to do with the ``mechanical stability''.

\subsection{Hydrogen atom: quantum mechanics}

The electron wave function $\phi$
of an hydrogen atom satisfies the the Schrodinger equation
\begin{align}\label{eq:schor}
    \left(-\frac{1}{2m}\nabla^2-eV_p(r)\right)\phi(\vec{r})=E\phi(\vec{r}) \ .
\end{align}
where ($e$ is the proton charge and positive)
\begin{align}
    \nabla^2 V_p(r)=-e\delta^3(\vec{r}) \ ,
\end{align}
is the potential of the static charge of the proton, $V_p=e/(4\pi r)$ . For convenient we chose $m=1$. One further defines the static potential $V_e$ for the electron
\begin{align}
\nabla^2 V_e(r)=e|\phi(\vec{r})|^2 \ ,
\end{align}
which can be solved for the ground state as
\begin{align}
    V_e(r)=\frac{e e^{-2\alpha r}(1+\alpha r-e^{2\alpha r})}{4\pi r } \ .
\end{align}
where $\alpha = e^2/4\pi$.

By non-relativistic reduction of the Dirac equation, one can construct the following EMT $T^{ij}_{\rm QM}$ which consists of a kinetic term
\begin{align}\label{eq:kine}
    T_{K}^{ij}=-\frac{1}{4m}\left(\phi^{\dagger}\partial^{i}\partial^{j}\phi-\partial^{i}\phi^{\dagger}\partial^{j}\phi+{\rm c.c}\right) \ ,
\end{align}
plus a potential term made of interacting electric fields of the proton and electron,
\begin{align}
    T_{V}^{ij}=\delta^{ij}\nabla V_p \cdot \nabla V_e- \partial^i V_e \partial^j V_p-\partial^iV_p\partial^jV_e \ .
\end{align}
The trace of $T_{\rm QM}^{ij}=T_{V}^{ij}+T_{K}^{ij}$ can be calculated as
\begin{align}\label{eq:TQM}
     T_{\rm QM}^{ii}=|\phi|^2(2E+eV_p)+\nabla\cdot (V_p\nabla V_e)+\frac{1}{4m}\nabla^2 |\phi|^2 \ .
\end{align}
It is easy to show that $T^{ij}_{\rm QM}$ is conserved for the ground state,  $\partial_i T^{ij}_{\rm QM}=0$.

Therefore, it can be written in a normalized state as
\begin{align}
    &T^{ij}_{\rm QM}(\vec{r})=(\delta^{ij}\nabla^2-\nabla^i\nabla^j)\frac{C_{\rm QM}(r)}{M} \ ,\\
    &T^{ii}_{\rm QM}(r)=2\nabla^2 \frac{C_{\rm QM}(r)}{M} \ ,
\end{align}
from which the $C_{\rm QM}$
can be calculated as
\begin{align}
    \frac{C_{\rm QM}(r)}{M}=\frac{1}{2\nabla^2}T^{ii}_{\rm QM}=\frac{e^{-2\alpha r} \alpha (2 \alpha r+1)}{16 \pi  r^2}-\frac{\alpha }{16\pi r^2} \ .
\end{align}
Clearly, the Coulomb tail $-{\alpha}/{16\pi r^2}$  prevents a finite $C(q=0)$.  Physically,
the self-energies of the proton and electron will generate opposite contributions which cancel
the Coulomb tail from the above expression (the EMTs
of the electron and proton are separately conserved). However,
this piece of physics is outside the usual non-relativistic quantum mechanics.

For the time being, we can subtract this
Coulomb tail and define an effective the $C_{\rm eff}(r)$ in the infared region where $r\sim \frac{1}{\alpha}$
\begin{align}\label{eq:Csimple}
\frac{C_{\rm eff}(r)}{M}=\frac{e^{-2 \alpha r} \alpha (2 \alpha r+1)}{16 \pi  r^2} \ ,
\end{align}
which is of order $1$ when the momentum transfer is of order of the inverse Bohr radius. In particular, the scalar MC monopole moment $\tau=\frac{C(0)}{M}$ for the hydrogen atom reads
\begin{align}\label{eq:hydroC0}
    \tau=\frac{C_{\rm eff}(q=0)}{M}=\tau_0[1 +{\cal O}(\alpha\ln \alpha)]\ ,
\end{align}
where $\tau_0=\frac{\hbar^2}{4m}$ is the basic unit defined before.  Below we show that in quantum field theory, the long range Coulomb tail is indeed removed and Eq.~(\ref{eq:hydroC0}) is the correct MC monopole moment.
\subsection{Hydrogen atom: field theory}
The above calculation can in fact be justified in the field theoretical framework. Let us consider the bound state in quantum electrodynamics (QED) between two types of fermions, the standard negative charged electron with mass $m$ and positive charged ``proton'' with mass $M$. At energy scale much smaller than the proton mass $M$, the proton can be approximated by an infinitely heavy static source $N$ represented by an auxiliary field $N$. The Lagrangian density of the system reads
\begin{align}
    {\cal L}=i\bar N v\cdot D N+{\cal L}_{\rm QED} \ ,
\end{align}
where $N$ represents the infinitely heavy proton moving along the $v^{\mu}=(1,0,0,0)$ direction. The Lagrangian preserves Lorentz invariance if $v^\mu$ is treated also as an auxiliary field. The EMT of the above system can be shown as
\begin{align}
    T^{\mu \nu}=\frac{i}{4}\bar N i v^{(\mu}D^{\nu)} N+T^{\mu \nu}_{\rm QED} \ .
\end{align}
More precisely, the heavy-source only contributes to the $T^{0i}$ part of the EMT.  To proceed, we fix the Coulomb gauge $\nabla \cdot \vec{A}=0$~\cite{Weinberg:1995mt}, in which the static potential $A^0$ can be solved as \begin{align}
    A^0(t,\vec{x})=-\frac{e}{\nabla^2}(\psi^{\dagger}\psi-\bar NN)(t,\vec{x}) \ .
\end{align}
By using $\vec{E}=-\partial_t \vec{A}_T-\nabla A_0$ and the explicit solution of $A_0$, the transverse and longitudinal parts of the electric field decouple from each other, and the Hamiltonian reads
\begin{align}
H&=\frac{1}{2}\int d^3\vec{x}\left(\vec{E}_T^2+\vec{B}_T^2\right)+\int d^3\vec{x}\psi^{\dagger}\left(-i\vec{\alpha}\cdot \vec{D}_T+m\gamma^0\right)\psi \nonumber \\
&+\frac{e^2}{2}\int d^3\vec{x}d^3\vec{y} \frac{(\psi^{\dagger}\psi-\bar N N)(\vec{x})(\psi^{\dagger}\psi-\bar N N)(\vec{y})}{4\pi |\vec{x}-\vec{y}|} \ ,
\end{align}
where the last term represents the Coulomb interaction.

\begin{figure}[t]
{%
  \includegraphics[height=2cm]{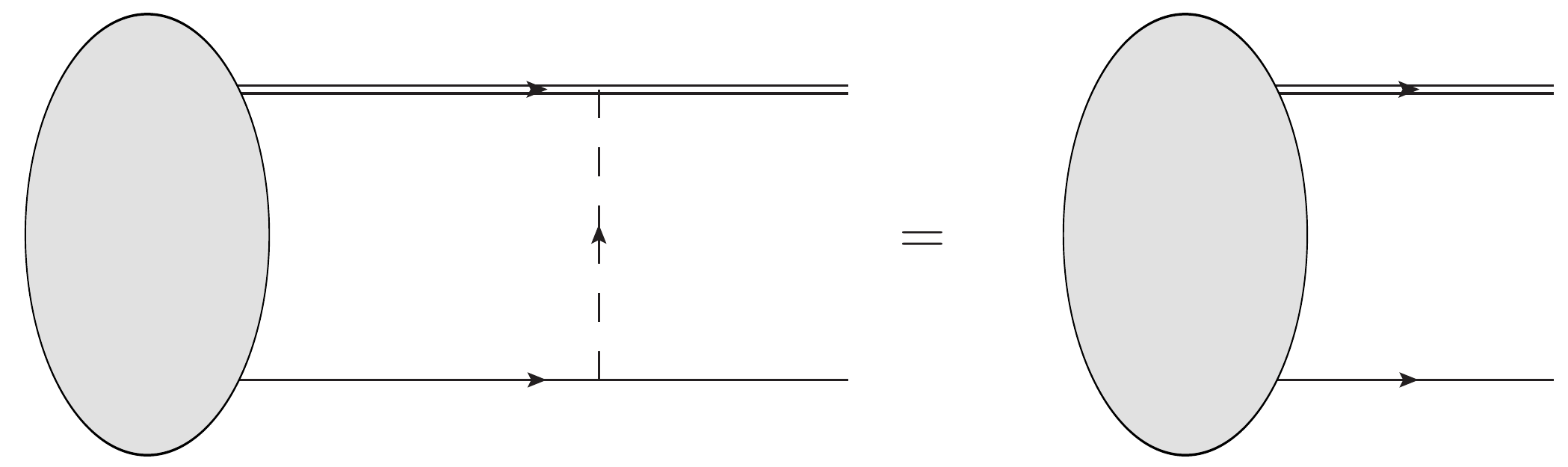}
}
\caption{The Bethe-Salpeter equation for the wave function $\phi$ denoted by the oval blob. Double line represents propagator of proton field and single line represents the electron propagator. The dashed line represents the exchange of a Coulomb photon. }
\label{fig:Bethe}
\end{figure}

We now consider the bound state formed by a pair of electron and the heavy-proton. The leading wave function reads
\begin{align}
    |\vec{p}\rangle=\int \frac{d^3\vec{k}}{(2\pi)^3}\frac{\phi(\vec{k})}{2E_{k}}|\vec{k}\rangle_{e}|-\vec{k}+\vec{p}\rangle_{N} \ ,
\end{align}
where to leading order in $\alpha$, $\phi(\vec{k})$ satisfies the Bethe-Salpeter equation~\cite{PhysRev.84.1232} induced by Coulomb-photon exchanges, see Fig.~\ref{fig:Bethe} for a depiction of the equation. It can be verified that it is nothing but the standard Schrodinger equation Eq.~(\ref{eq:schor}) in non-relativistic limit $|\vec{k}|\ll m$, and is characterized by two scales, the binding energy $\alpha^2m$ and the inverse Bohr radius $\alpha m$.

\begin{figure}
{%
  \includegraphics[height=4cm]{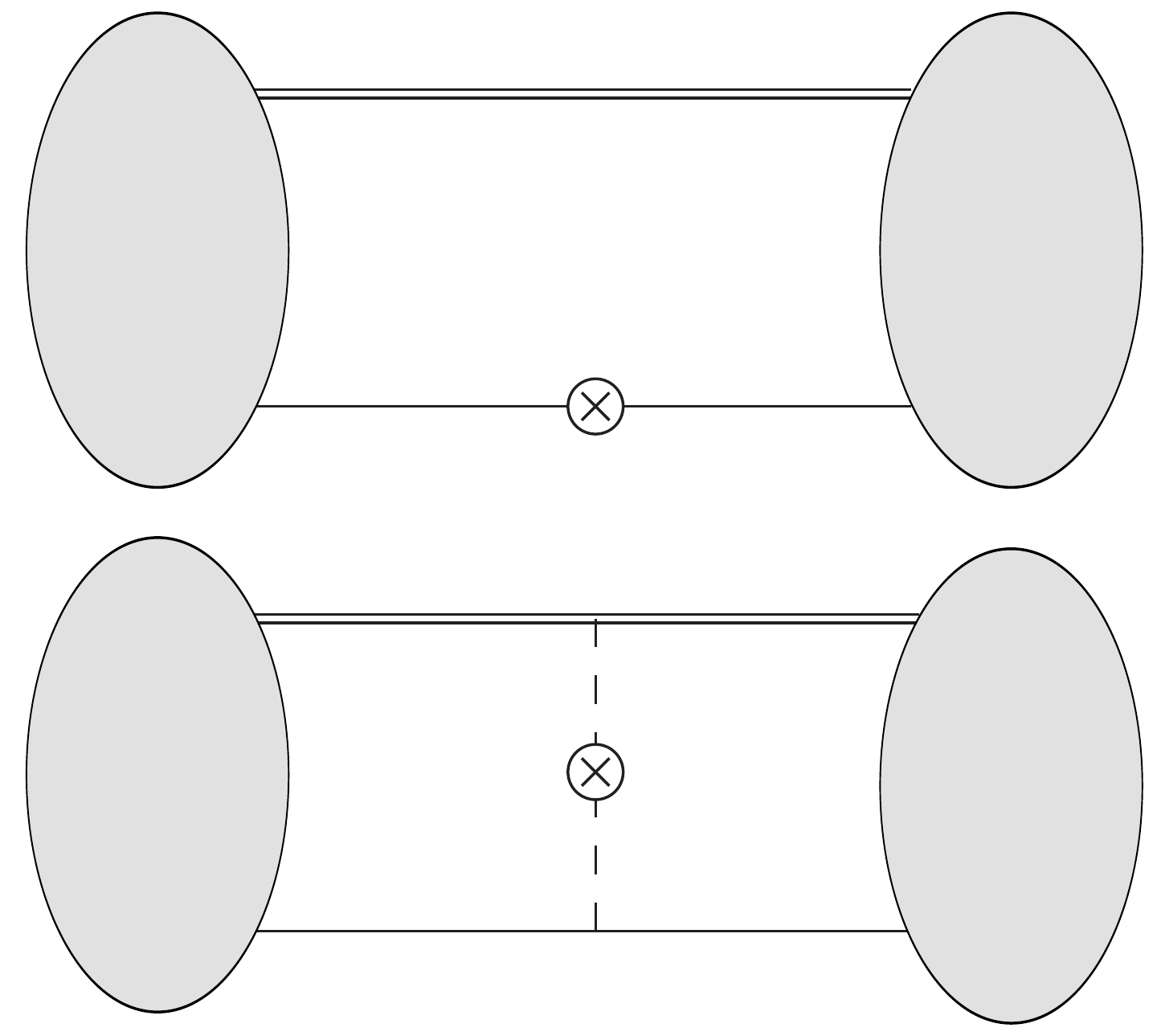}
  \includegraphics[height=1.98cm]{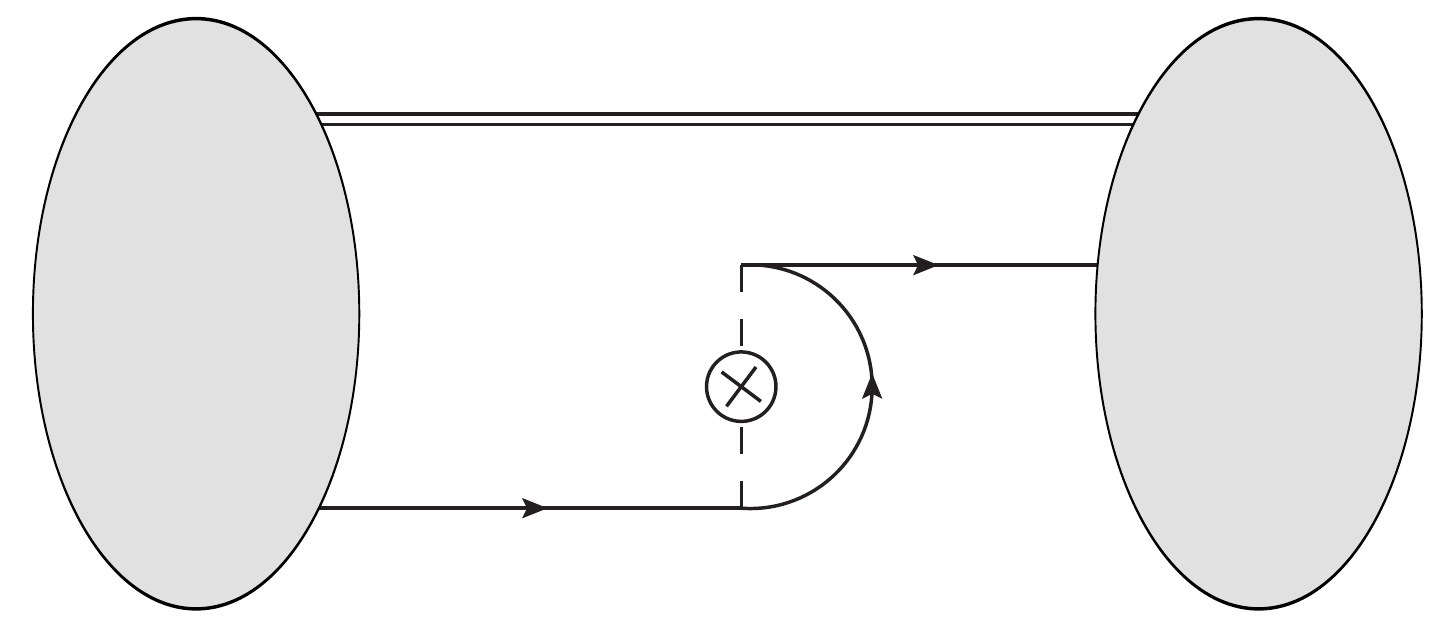}
}
\caption{The order-${\cal O}(1)$ electron kinetic contribution (upper), the Coulomb photon interference (middle) and single electron (lower) contributions to $T^{ij}$. Dashed lines represent Coulomb photons and crossed circles denote the operator insertions. Notice the infrared divergences for $C(q)$ at $q=0$ are cancelled between the interference and single electron and proton (not shown) contributions.}
\label{fig:inter}
\end{figure}

\begin{figure}[t]
{%
  \includegraphics[height=5cm]{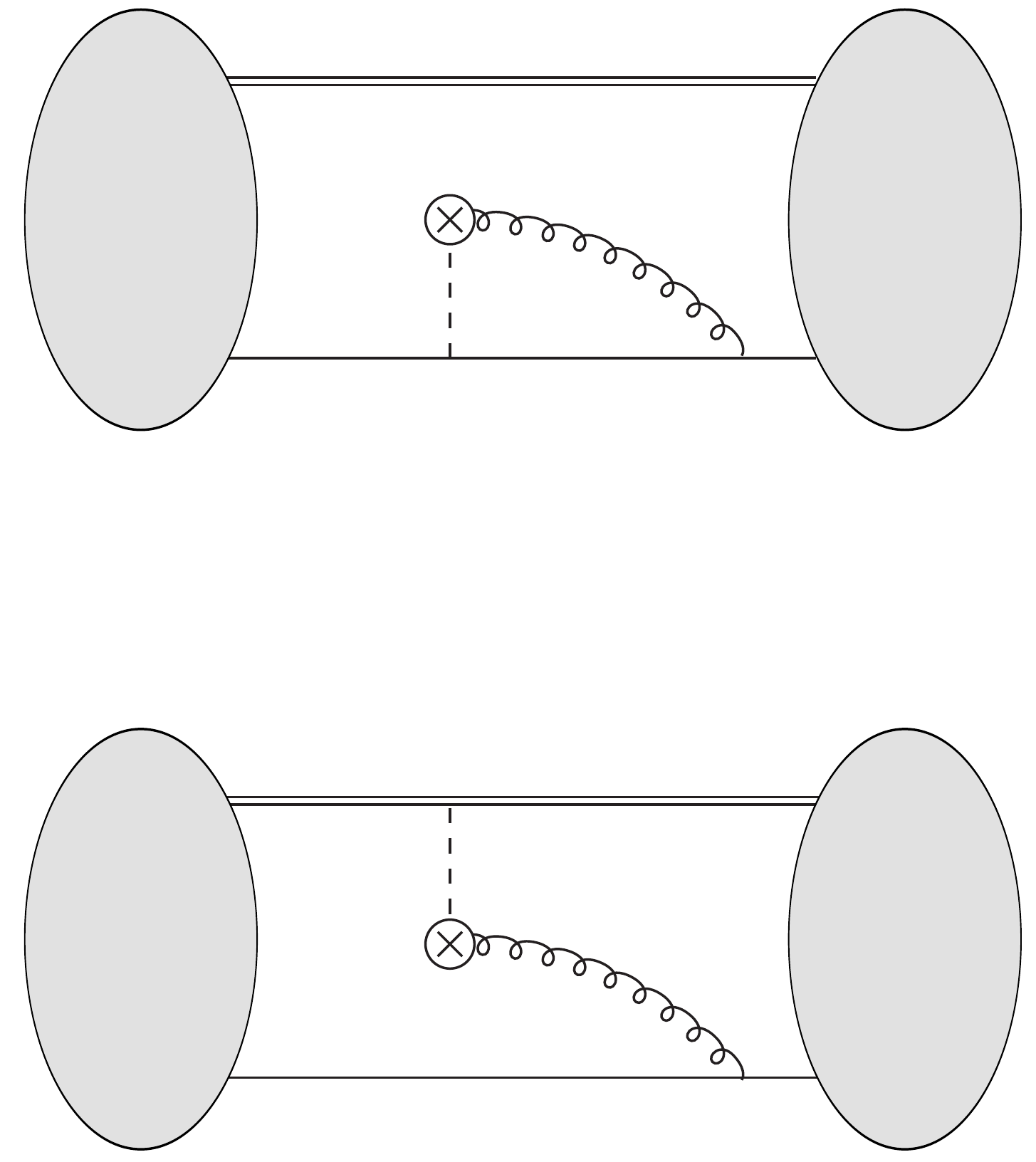}
}
\caption{Mixed contributions between the radiative (wavy line) and Coulomb photons. IR divergences are regulated by binding energy differences. }
\label{fig:lamb}
\end{figure}

Given the above wave function, one can calculate the matrix element of $T^{ii}(\vec{r})$ using standard method. At leading order ${\cal O}(1)$ in $\alpha$, there is only one diagram, the electron kinetic  contribution shown in Fig.~\ref{fig:inter}. One can show that this contribution is exactly the $T_{\rm K}^{ij}$ in Eq.~(\ref{eq:kine}).

We then consider radiative corrections starting from ${\cal O}(\alpha)$. Due to the fact that the velocities are of order $\alpha$, only the Coulomb-photon contribution needs to be included. Furthermore, the one-loop contributions can be classified into interference and single electron/single proton diagrams, see Fig.~\ref{fig:inter} for a depiction. The interference diagram can be calculated as
\begin{align}
T^{ii}(\vec{q})= -e^2 \int \frac{d^3kd^3k'}{(2\pi)^6}\phi^{\dagger}(k')\phi(k-k')\frac{\vec{k}\cdot (\vec{k}-\vec{q})}{\vec{k}^2(\vec{k}-\vec{q})^2} \ ,
\end{align}
which is equivalent to
\begin{align}
  T^{ii}_V=  \nabla V_e \cdot \nabla V_p  \ .
\end{align}
Therefore, we found that the leading-order electron kinetic and interference contributions in Fig.~\ref{fig:inter} will exactly lead to the conserved $T_{QM}^{ii}$ in Eq.~(\ref{eq:TQM}) with a Coulomb tail $\frac{\pi \alpha}{8|q|}$.

Clearly, in order to cancel the Coulomb tail one must add the electron and proton self-energy contributions. The electron contribution is also shown in Fig.~\ref{fig:inter} and can be calculated in the region $q \sim {\cal O}(\alpha m)$ as
\begin{align}
   \frac{ C_{\rm e} (q)}{M}= \frac{\alpha \pi}{16|q|}\times \frac{16\alpha^4}{(\frac{q^2}{m^2}+4\alpha^2)^2} \ ,
\end{align}
where the first factor $\frac{\alpha \pi}{16|q|}$ is just the standard Coulomb tail, and the second factor comes from the dressing in the bound-state wave function. Similarly, the proton contribution is of the form in the $M \gg m$ limit
\begin{align}
    \frac{C_{\rm p} (p)}{M} = \frac{\alpha \pi}{16|q|}\times \frac{16\alpha^4}{(\frac{q^2}{M^2}+4\alpha^2)^2} \ ,
\end{align}
and can be approximated simply by the Coulomb tail when $q\sim \alpha m$. Clearly, the above formulas can be justified only when the momentum transfer is small, and is invalid in ultraviolet region when $q$ is comparable with particle masses.

In conclusion, in the region $|q|\le{\cal O} (\alpha m)$, the $C$ form factor of the hydrogen atom reads
\begin{align}\label{eq:hydrogenfinal}
    \frac{C_{\rm H}(q)}{M}&=\frac{1}{2m(\frac{q^2}{\alpha^2m^2}+4)}-\frac{\alpha}{4|q|}\left(\frac{\pi}{2}-{\rm Arctan} \frac{q}{2\alpha m}\right) \nonumber \\
    &+\frac{\alpha \pi}{|q|}\frac{1}{(\frac{q^2}{\alpha^2m^2}+4)^2} +\frac{\alpha \pi}{|q|} \frac{1}{(\frac{q^2}{\alpha M^2}+4)^2} \ .
\end{align}
From these, the monopole moment  for the hydrogen atom is
\begin{align}
    \tau_{\rm H}=\frac{C_{\rm H}(0)}{M}=\tau_0[1 +{\cal O}(\alpha \ln \alpha)]\ .
\end{align}
The $g_H=1$ except for a small correction of order $\alpha$, a result with opposite sign from a point-like boson.

\begin{figure}[t]
{%
  \includegraphics[height=5.5cm]{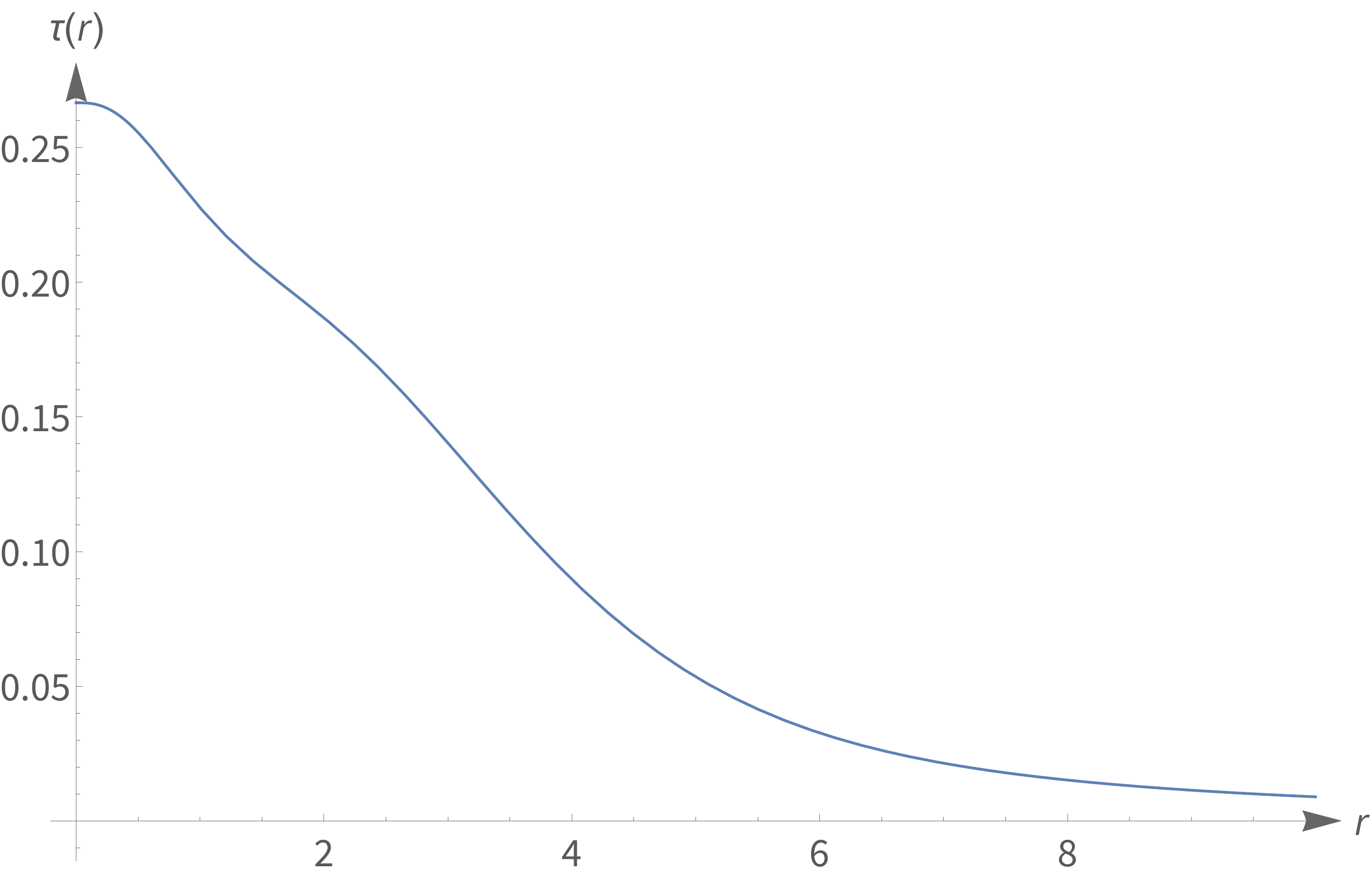}
}
\caption{The tensor momentum-current monopole density $\tau(r)$ in Eq.~(\ref{eq:taur}) as a function of $r$ in hydrogen atom, where $r$ is in unit of the Bohr radius $a_0=\frac{\hbar}{\alpha mc}$ and $\tau(r)$ is in unit of  $\frac{\hbar^2}{4m}\frac{1}{a_0}$ with SI dimension $\frac{{\rm kg}\cdot {\rm m}^3}{{\rm s}^2}$. It is positive and finite as $r \rightarrow 0$. The proton mass $M$ has been approximated by $\infty$. }
\label{fig:taur}
\end{figure}

\begin{figure}[t]
{%
  \includegraphics[height=6cm]{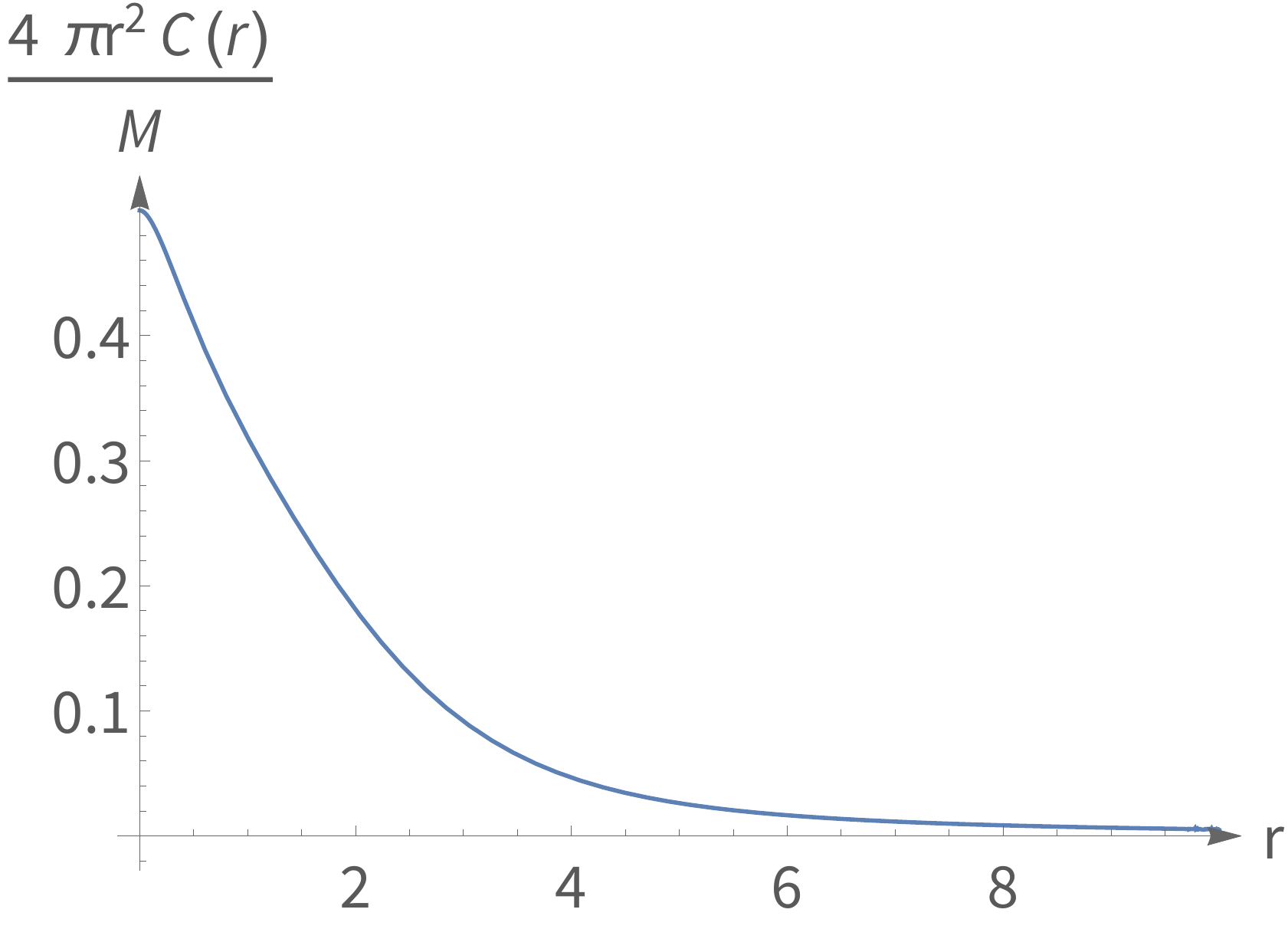}
}
\caption{The $4\pi r^2\frac{C(r)}{M}$ as a function of $r$ in hydrogen atom, where $r$ is in unit of the Bohr radius $a_0=\frac{\hbar}{\alpha mc}$ and $4\pi r^2\frac{C(r)}{M}$ is in unit of  $\frac{\hbar^2}{4m}\frac{1}{a_0}$. It has the same SI dimension $\frac{{\rm kg}\cdot {\rm m}^3}{{\rm s}^2}$ as that of $\tau(r)$. It is positive and finite as $r \rightarrow 0$. According to Eq.~(\ref{eq:hC}), it contributes to a part of the gravitational potential inside the hydrogen atom.}
\label{fig:Cr}
\end{figure}

Given the form factor $C(q^2)$, we can obtain the scalar momentum current or  ``pressure"~\cite{Polyakov:2018zvc,Lorce:2018egm},
\begin{align}
    p(r)=\frac{1}{3}T^{ii}= \frac{2}{3M}\frac{1}{r^2}\frac{d}{dr} r^2\frac{dC(r)}{dr} \ ,
\end{align}
which can be shown to be positive for small $r$ and negative for large $r$. On the other hand, the momentum current monopole density distribution
\begin{equation} \label{eq:taur}
   \tau(r) = -\frac{2\pi}{5}r^2
        \left(r^ir^j-\frac{1}{3}r^2\delta^{ij}\right)
        T_{ij}(r) = -\frac{4\pi}{15}r^4s(r)
        \ .
\end{equation}
where $s(r)=-\frac{r}{M}\frac{d}{dr}(\frac{1}{r}\frac{dC(r)}{dr})$ has been called ``shear pressure''.
 Unfortunately, there is no simple expression for it in positional space without using the Meijer $G$ function. A numerical result for $\tau(r)$ is shown in Fig.~\ref{fig:taur}, which is positive definite at all $r$ and finite at $r=0$. A plot of $C(r)$ is shown in Fig.~\ref{fig:Cr}.

\begin{figure}[t]
{%
  \includegraphics[height=8cm]{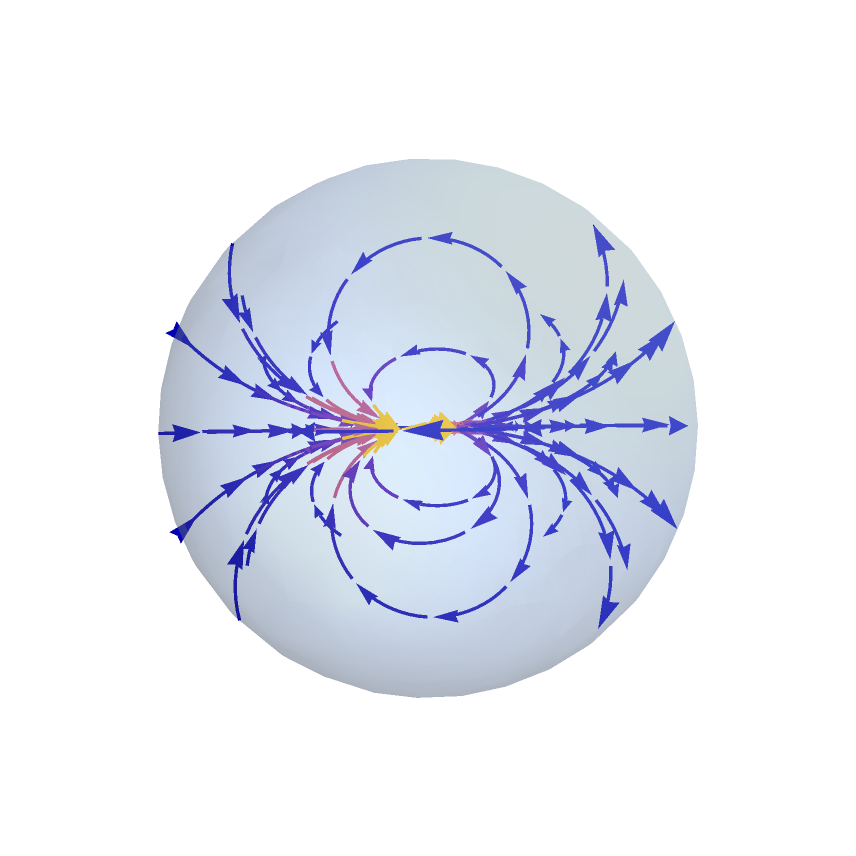}
}
\caption{The conserved $x$-component-momentum current distribution $T^{ix}(\vec{r})$ as a vector field in $i$, with the arrows indicating directions of the current flow. The $x$-direction points to the right in the horizontal direction.  }
\label{fig:flow}
\end{figure}

We also show a plot of the momentum flow $T^{ix}$ in 3-space in Fig.~\ref{fig:flow}.
Any surface integral of the vector  field  yields  the  flux  of $x$-component  momentum through the surface, or force along the $x$-direction, which would produce a pressure in this direction if the momentum current gets totally absorbed.  Any closed surface integral yields zero, indicating momentum conservation or net null force through any size volume.

Before ending the section, one briefly comment on the ${\cal O}(\alpha)$ corrections to $C(0)$. By including radiative photons, the degree of infrared divergences is reduced due to the fact that the radiative photon couples to the $3$-velocity $\vec{v}$ of the electron in the Coulomb gauge. Therefore, the mixing diagram where Coulomb and radiative photons couple to each other can be logarithmically divergent, confirmed by the single electron calculation~\cite{Berends:1975ah,Milton:1976jr,Donoghue:2001qc,Varma:2020crx,Metz:2021lqv}. However, the divergences is expected to be regulated by the binding energy differences in a way similar to the famous Lamb shift, leading to finite $C(0)$ at order ${\cal O}(\alpha)$. See Fig.~\ref{fig:lamb} for a depiction.
We will leave the calculation of order $\alpha$ corrections in a separate publication~\cite{future}.

\section{Comments on Mechanical Stability and Conclusions}

The QCD EMT of a hadron has often been related to
the stress tensor of a continuous medium~\cite{Polyakov:2002wz}. If
this analogy is of value, one can introduce
the concept of pressure, shear pressure,
mechanical stability, etc.\cite{Perevalova:2016dln,Lorce:2017xzd, Polyakov:2018zvc,Lorce:2018egm}, and try to
understand a bound state in quantum theory using
the terms of classical mechanics. However, this analogy is of limited use and can even be misleading at times.

The scalar momentum-current $T^{ii}(r)$ has been identified as the pressure $p(r)$, which is true in some models of continuous media.  However, the two concepts have different physical significance and are not interchangeable in general. The concept of pressure normally stands for an isotropic force
from random microscopic motions in all possible directions and is a positive quantity for stable systems. However, $T^{ii}(r)$ is not positive definite, and carries with it the sense of a directional flow, analogue of {\it macroscopic motion} in a fluid.

The proper analogue of the momentum currents $T^{ij}$ may be the radiation pressure~\cite{Jackson:1998nia,Noguchi2020}: Assuming a directional momentum current gets absorbed on a surface, the surface experiences a force or a pressure that can be measured by the momentum passing through the surface per unit time. In this case, the pressure is not a scalar as in thermal systems, rather it depends on the directions of the momentum flow as well as the surface area. Thus a negative pressure only means the momentum flow is negative with respect to a reference direction. The Laue condition
\begin{equation}
     \int dr~r^2 p(r)= 0 \ ,
\end{equation}
is trivially related to the conservation of
the momentum current~\cite{Laue:1911lrk, Polyakov:2018zvc}.
Using ``pressure'' or ``force'' to characterize the momentum current may generate confusion because they are not acting on any part of the system itself, but on some fictitious surfaces which would absorb the current entirely through some  interactions. In our view,
the most interesting way to characterize $T^{ii}$ is by its multipoles generating characteristic space-time perturbation. This is similar to use the magnetic moment etc.
to describe a current distribution, which generates a particular type of magnetic field.

Further mechanical stability conditions on $T^{ij}$ derive from comparing it the pressure and shear pressure distributions with a mechanical system. For example, it has been speculated that a negative $D$-term ($C(q=0)$) is needed for mechanical stability~\cite{Perevalova:2016dln}.
However, the sign of the scalar-MC monopole moment depends the flow pattern of the momentum currents. Reversing the direction of the momentum currents at every point in space in classical physics will reverse the $D$-term but should lead to another stable flow pattern. Thus the sign of $D$-term cannot be related to mechanical stability.
In fact, in the example of hydrogen atom, the ``force'' from the momentum flow is directed towards the center (Using the $C_{\rm eff}(r)$ defined in Eq.~(\ref{eq:Csimple})) and the $D$-term is positive. On the other hand, we know perfectly well that hydrogen atom is stable due to quantum physics which has already a {\it well-defined sense of  stability}. Thus using
the momentum current flow to judge the stability of a quantum system appears not useful and furthermore unnecessary.

To summarize, we have revisited the gravitational fields generated by a static source by performing the multipole expansion~\cite{Damour:1990gj,Maggiore} of the corresponding energy, energy current and momentum current densities. For a static and conserved EMT, there are $6$ series of non-vanishing multipoles, one series for $T^{00}$, two series for $T^{0i}$ and three series for $T^{ij}$. They are important characterization of spatial distribution of the corresponding energy and momentum current distributions, although at large distance only three series will contribute to the physical metric perturbation in the static case.

In particular, the $C(q^2)$ form factor or the ``$D$-term'' is related to the tensor-MC monopole moment $T0$, which has a basic unit $\tau_0=\hbar^2/4M$. As a concrete example, we have calculated the $C(q^2)$ form factor for the hydrogen atom at small $q$ region and found that the MC monopole moment is positive, which is opposite to that of a point-like boson. Moreover, we argue that notion of ``mechanical stability" or ``presure'' is of limited significance when applied to bound states in quantum field theories.

{\it Acknowledgment.}---We thank A. Buonanno, V. Burkert, T. Damour, L. Elouadrhiri, F.-X.Girod, K.-F. Liu, P. Schweitzer, and T. Jacobson for valuable discussions and correspondences, for Yuxun Guo for drawing Fig. 4, and for E. Jiang to participate in the early calculation for the hydrogen atom.
This research is supported by the U.S. Department of Energy, Office of Science, Office of Nuclear Physics, under contract number DE-SC0020682, and the Center for Nuclear Femtography, Southeastern Universities Research Association, Washington D.C.

\bibliography{bibliography}

\end{document}